# ARTICLE

# What's special about Y6; the working mechanism of neat Y6 organic solar cell

Elifnaz Sağlamkaya[a], Artem Musiienko[b], Mohammad Saeed Shadabroo[a], Bowen Sun[a], Sreelakshmi Chandrabose[c], Giulia Lo Gerfo M.[d], Niek F. van Hulst[d], Dieter Neher[c], Safa Shoaee*[a]

Non-fullerene acceptors (NFA) have delivered advance in bulk heterojunction organic solar cell efficiencies, with the significant milestone of 20% now in sight. However, these materials challenge the accepted wisdom of how organic solar cells work. In this work we present neat Y6 device with efficiency above 4.5%. We thoroughly investigate mechanisms of charge generation and recombination as well as transport in order to understand what is special about Y6. Our data suggest Y6 generates bulk free charges, with ambipolar mobility, which can be extracted in the presence of transport layers.

## Introduction

Learning from the success of the fullerene acceptors, research groups synthesized non-fullerene acceptors (NFAs) from 2D planar molecules into "3D-like" molecules. With this strategy in mind, the emergence of the Y-series NFAs has dramatically pushed organic solar cells (OSC) forward. Highest efficiencies are reported for single junction of ternary blends in which one of the components is the NFA (2,2'-((2Z,2'Z)-((12,13-bis(2-ethylhexyl)-3,9-diundecyl-12,13-dihydro-[1,2,5]thiadiazolo[3,4-e]thieno[2",3'':4',5']thieno[2',3':4,5]pyrrolo[3,2-g]thieno[2',3':4,5]thieno[3,2-b]indole-2,10-diyl)bis(methanylylidene))bis(5,6-difluoro-3-oxo-2,3-dihydro-1H-indene-2,1-diylidene))dimalononitrile), Y6 acceptor, or one of its close derivatives. Combining Y6 with PM6 gives high and reproducible power conversion efficiencies (PCEs), between 15% to 19%.[1,2,3,4,5] Surprisingly high external quantum efficiencies (EQEs) are observed even in the absence of a considerable driving force.[4,6,7] In fact, in PM6:Y6 free charge generation was shown to be essentially barrierless.[4] One of the puzzles is the efficient exciton dissociation efficiency with respect to the simple Coulomb-based arguments.

When light is absorbed by organic semiconductors, the tenet holds that only excitons are intrinsically photogenerated. The excitons persist due to the strong Coulombic interaction between electrons and holes, leading to large (200 – 1000 meV)[8] exciton binding energies ($E_b$). This has led researchers to combine two materials with large energetic offset, to overcome the $E_b$. Such paradigms have guided the development of OSCs for the last three decades. However, the small energetic offset between Y6 and several polymers with yet high efficiencies,[9] has sparked a lot of interest in understanding the mechanism of charge generation in neat Y6. The demand of driving force for exciton dissociation in OSCs is ascribed to the strong $E_b$.[10,11] Therefore, the success of high efficiencies with a small energetic offset makes Y6 intriguing as single-component materials and calls for a re-evaluation of charge generation mechanisms in new materials and systems.

Formation of polaron pairs in neat Y6 films has now been reported by several groups.[12,13] Transient absorption studies have suggested that in neat Y6, excitons are delocalised, or form an intra-moiety intermediate state.[13] Interestingly, all existing time-resolved spectroscopic studies indicate that both excitons and polaron pairs are simultaneously created within the time resolution of the experiment,[14] however since both features inconveniently overlap in the spectral region, and the free charges are reported to be very short-lived, it is challenging to understand the mechanism of charge generation. This puzzling observation has led to a substantial debate about the physical mechanisms underlying ultrafast polaron pair formation in neat Y6. Different scenarios can be invoked, including field-assisted photogeneration of polaron pairs, the rapid formation

[a.] *Disordered Semiconductor Optoelectronics, Institute of Physics and Astronomy, University of Potsdam, Karl-Liebknecht-Str. 24-25, 14476 Potsdam-Golm, Germany. E-mail: shoai@uni-potsdam.de*

[b.] *Department Novel Materials and Interfaces for Photovoltaic Solar Cells, Helmholtz-Zentrum Berlin für Materialien und Energie, Kekuléstraße 5, 12489 Berlin, Germany*

[c.] *Soft Matter Physics, Institute of Physics and Astronomy, University of Potsdam, Karl-Liebknecht-Str. 24-25, 14476 Potsdam-Golm, Germany*

[d.] *ICFO – Institut de Ciencies Fotoniques, The Barcelona Institute of Science and Technology, 08860 Castelldefels, Barcelona, Spain.*

† Electronic Supplementary Information (ESI) available: [details of any supplementary information available should be included here]. See DOI: https://doi.org/10.1039/D2MH01411D

of polaron pairs by exciton dissociation either at interlayer interfaces or due to energetic cascade caused by aggregation and different crystallinity and morphology of Y6 or, polaron pair formation from delocalized coherent excitations due to low $E_b$. In this regard Zhu et al. have estimated the $E_b$ of Y6 in solid phase to yield values of 0.11–0.15 eV.[15] Such small values are debatable compared to cyclic voltammetry measurements where exciton binding energy of around 0.3 eV can be estimated.[16] Nevertheless, all of these values are larger than thermal energy, $k_BT$, which make spontaneous efficient charge generation implausible. In addition, another drawback of neat devices is the difficulty obtaining the optimized phase separation between donor (D) and acceptor (A) segments in a single material, resulting in severe charge recombination and low charge transport efficiency, as reported by Hodgkiss and co-workers.[12]

To reflect the importance of all of these exhilarating findings, we fabricate single component Y6 devices with an efficiency of 4.5% - the highest PCE reported for a single material small molecule OSCs.[17,18,19] We use a combination of electroluminescence quantum efficiency ($EQE_{EL}$), photoluminescence (PL), electroluminescence (EL), temperature dependent quasi-steady state absorption spectroscopy (PIA), Hall and photo-Hall measurements, time-delayed collection field (TDCF), and bias-assisted charge extraction (BACE) to analyse the device physics and identify mechanism of charge generation and recombination pathways. Whilst exciton dissociation has been reported on ultrafast timescales,[12] herein we observe that long-lived charge is only present when Y6 is interfaced with at least one transport layer. Our Hall measurements, PIA and TDCF data indicate long lived free charges are generated in the bulk and are field independent, whilst the field dependent PL measurements further confirm extraction of free charge. We further find that charge transport is bipolar and very efficient: $2 \times 10^{-3}\ cm^2/Vs$ and $0.9 \times 10^{-3}\ cm^2/Vs$ for electrons and holes respectively. However, despite these merits, recombination of free carriers in neat Y6 device is up to one order of magnitude faster than the state-of-the-art PM6:Y6 blend. We also investigate single neat devices of (2,2'-((2Z,2'Z)-((12,13-bis(2-ethylhexyl)-3,9-diundecyl-12,13-dihydro[1,2,5]thiadiazolo[3,4e]thieno[2'',3'':4',5'] thieno[2',3':4,5]pyrrolo[3,2-g] thieno[2',3':4,5]thieno[3,2-b]indole-2,10-diyl)bis(methanylylidene))bis(3-oxo-2,3-dihydro1H-indene-2,1-diylidene))dimalononitrile), Y5, and (2,2'-((2Z,2'Z)-((12,13-bis(2-butyloctyl)-3,9-dinonyl-12,13-dihydro-[1,2,5]thiadiazolo[3,4-e]thieno[2'',3'':4',5']thieno[2',3':4,5]pyrrolo[3,2-g]thieno[2',3':4,5]thieno[3,2-b]indole-2,10-diyl)bis(methanylylidene))bis(5,6-dichloro-3-oxo-2,3-dihydro-1H-indene-2,1-diylidene))dimalononitrile), eC-9 for comparison. From the findings we suggest that the given external quantum efficiency (EQE) in Y6 is due to bulk charge generation in Y6, good charge transport as well as the presence of transport layers to spatially separate the free carriers from one another.

## Results and discussion

In the investigated devices, Y6 is inserted between ITO and Al contacts, together with electron and hole transport layers (ETL and HTL, respectively), which improve the contact selectivity and device performance. 3-[6-(diphenylphosphinyl)-2-naphthalenyl]-1,10-Phenanthroline (Phen-NaDPO or DPO) is used as ETL, and Copper(I) thiocyanate (CuSCN) is used as HTL. As shown in **Figure 1a**, 60 nm layer of CuSCN is used in adjacent to Y6, as hole transport layer whilst CuSCN does not absorb to generate any charges.[14] Further details about the device structure can be found in the Experimental Section. The best device performance is achieved when adding 0.5 % 1, 8-diiodooctane (DIO) to chloroform ($CHCl_3$) as additive to the Y6 solution. The open-circuit voltage ($V_{OC}$) in both cases is rather low as compared to Y6's optical gap ($E_{opt}$) of 1.7 eV.[20] The biggest difference with the addition of the additive is reflected in the increase in the average short-circuit current density ($J_{SC}$) from 5.3 to 8.4 mAcm$^{-2}$ and in the fill-factor (FF) 65 % compared to 55 % (**Figure 1b**). Yet interestingly we do not observe any sub-gap absorption peaks in the measured EQE spectra which could be attributed to the formation of charge transfer (CT) states at the Y6 and interlayer interfaces (Figure S1a, Supporting Information) and the thickness dependent data shows an increase in $J_{SC}$ and performance with increasing junction thickness (Figure S2). In this paper we focus to understand the mechanism of charge generation with and without DIO, reflecting on the $J_{SC}$ and FF.

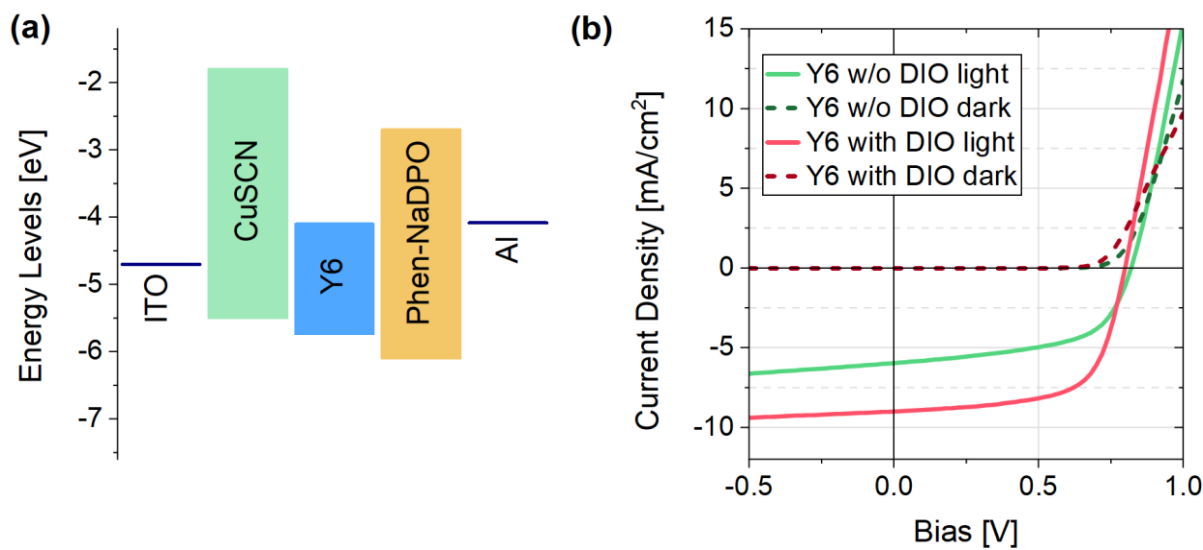

**Figure 1a.** Energy levels (materials not in contact with each other) of CuSCN[21] hole transport layer, small molecule acceptor Y6,[22] Phen-NaDPO[23] electron transport layer and of the electrodes[21] taken from the literature. **b.** Current density–voltage (*JV*) characteristics of a regular device with a 100 nm neat Y6 active layer with and without DIO additive measured under simulated AM1.5G light (solid lines) and in the dark (dashed lines).

| Additive | $J_{SC}$ [mA cm$^{-2}$] | $V_{OC}$ [V] | FF [%] | PCE [%] | $\Delta V_{OC, nr}$ |
|---|---|---|---|---|---|
| with DIO | 8.4 | 0.79 | 65 | 4.4 | 0.3 |
| without DIO | 5.3 | 0.82 | 55 | 2.4 | 0.26 |

**Table 1** Average device characteristics for neat Y6 devices under AM1.5 illumination for a minimum of 5 individual devices, and voltage losses for the best performing device calculated from the electroluminescence quantum yield (ELQY). For clarity, the errors are shown in Table S2, Supporting Information.

To elucidate the free charge separation mechanism in the neat Y6 device, we employed Hall effect and photoconductivity measurements on 100 nm thick neat Y6 film on glass with and w/o DIO. Photo-Hall measurements are made by monitoring the Hall signal with increasing illumination intensity. The rise in the photoconductivity indicates the formation of the free charge carriers. The charge carrier concentration can be calculated form the product of the photoconductivity and the mobility (see supplementary note 1.) In the dark measurements, we calculated the carrier density with the assumption that the Hall signal originates from the excess electrons, in this case, the neat Y6 films. The films w/o DIO had carrier density of $3.8 \times 10^{13}$ cm$^{-3}$. The solvent additive DIO introduces a major increase in the electron density estimated to be $7.4 \times 10^{14}$ cm$^{-3}$. Under illumination with the excitation energy of 1.7 eV, both films exhibited an increase in the photoconductivity due to the increased carrier concentration, which points towards bulk charge photogeneration in neat Y6 films on glass (**Figure 2a**).

To assess the efficiency of free carriers in films and with transport layers, we employ quasi steady-state photoinduced absorption (PIA) spectroscopy. We have previously demonstrated that PIA can be an effective assay of yield of free carriers.[24] The technique monitors the differential absorption upon modulation of the intensity of the quasi steady-state illumination. **Figure 2b** shows the PIA spectra measured on CuSCN/Y6/DPO sample under open-circuit conditions. For Y6 films sandwiched between the interlayers we observe the GSB signal around 850 nm and the PIA band in the region around 1000 nm. Given that the technique operates at long time scale measurements ($\mu$s-s), unique absorption features of polarons are of those of free carriers and not charge transfer states or excitons, indicating that long-lived charges are present in the sample. In particular its observed that the DIO sample exhibits a much higher amplitude. The amplitude of the signal is proportional to carrier density, *n* through the equation $\frac{\Delta T}{T} = \sigma \times n \times d$;[24] thus insinuating the DIO sample has much higher free carriers. On the other hand, films of Y6 without transport layers gave negligible signals. This highlights the important role of the transport layers in spatially separating and stabilising the free carriers from the photoactive layer.

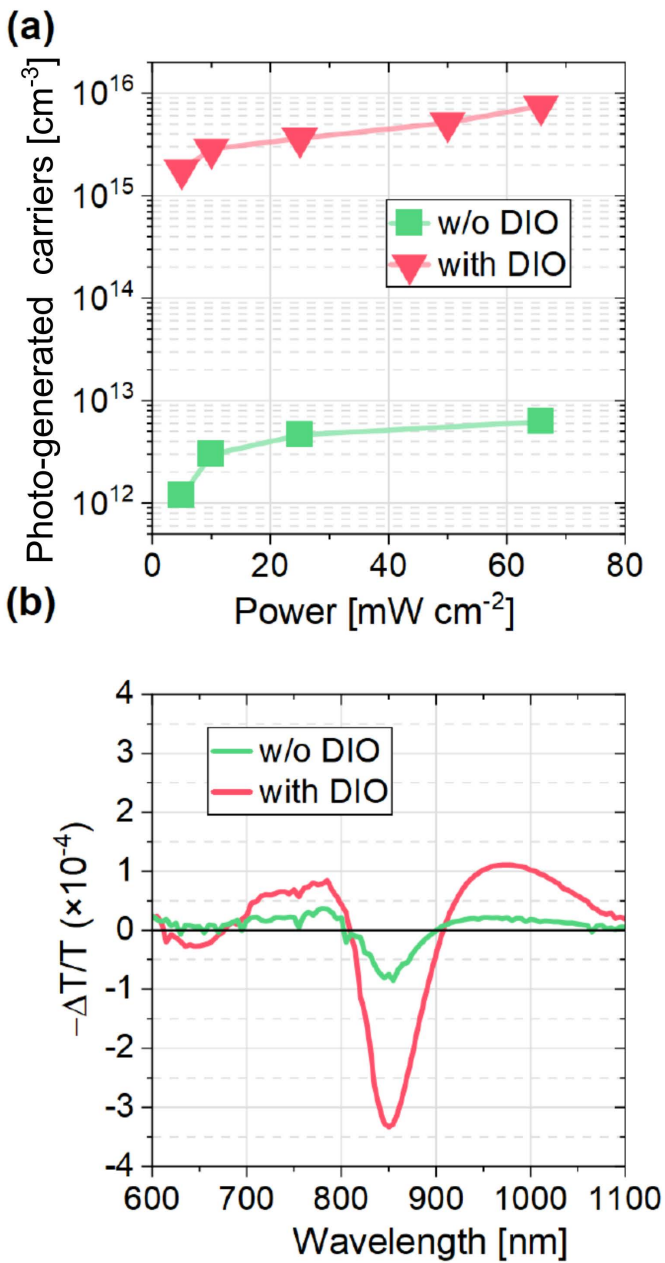

**Figure 2a.** The Photo-generated carriers in the 100 nm thick neat Y6 films on glass was estimated from intensity dependent 4-probe Hall conductivity measurement with the excitation energy of 1.7 eV **b.** Photo-induced absorption spectra of the semi-transparent with and w/o Y6 devices measured from the electrode-free area.

To understand the mechanism of charge generation in neat Y6 and the difference in the PIA amplitudes between DIO and no DIO samples, we performed photoluminescence (PL) measurements to elucidate if excitons limit charge generation. In order to investigate the role of excitons, we fabricate thin Y6 (15 nm) films with and without the transport layers, to measure photoluminescence quantum yield (PLQY) and thereby assess exciton quenching As shown in **Figure 3a**, the PL of neat Y6 film prepared from $CHCl_3$ only, gets quenched by 76 % when interfaced with both CuSCN and Phen-NaDPO layers, whilst when Y6 is processed from $CHCl_3$ and DIO, and sandwiched between CuSCN and Phen-NaDPO, we measure Y6's singlet exciton to be quenched by 98 % (see Figure S5 in for thickness dependent PLQ data). Spatially and time resolved fluorescence microscopy measurements conducted on DIO and no DIO samples, indicates that exciton diffusion coefficient (D) of the 100 nm Y6 film with DIO is only higher by 30% than the D of the Y6 (w/o DIO) while having shorter singlet lifetime. The calculated exciton diffusion length of the Y6 (with DIO) film is only 7 nm longer than Y6 (w/o DIO) which doesn't explain the large difference in the $J_{SC}$ between the two devices (see supplementary note 2). Consistent with reports on aggregation of Y6,[25,26] our result (in particular the thickness dependent JV data in Figure S2) favour a picture of aggregation-dependent energy levels,[27] thereby enabling an energy cascade which facilitates to dissociate excitons and further drive charge out from the more disordered domains into a more-crystalline domains of Y6. But it should be noted that whilst we see quenching of Y6 emission when interfaced with transport layers, we cannot rule out partial quenching in the neat Y6 film on glass. The enhanced PL quenching of the DIO sample compared with no DIO just on glass, indicates that some exciton dissociation readily occurs in the neat film on glass without the aid of transport layers. Furthermore, very strikingly, complete devices under different bias conditions ($V_{OC}$, $J_{SC}$ and reverse bias) found to have decreasing PL intensity with increasing reverse bias (**Figure 3b**) indicating that PL intensity in neat Y6 device is a field assisted process.

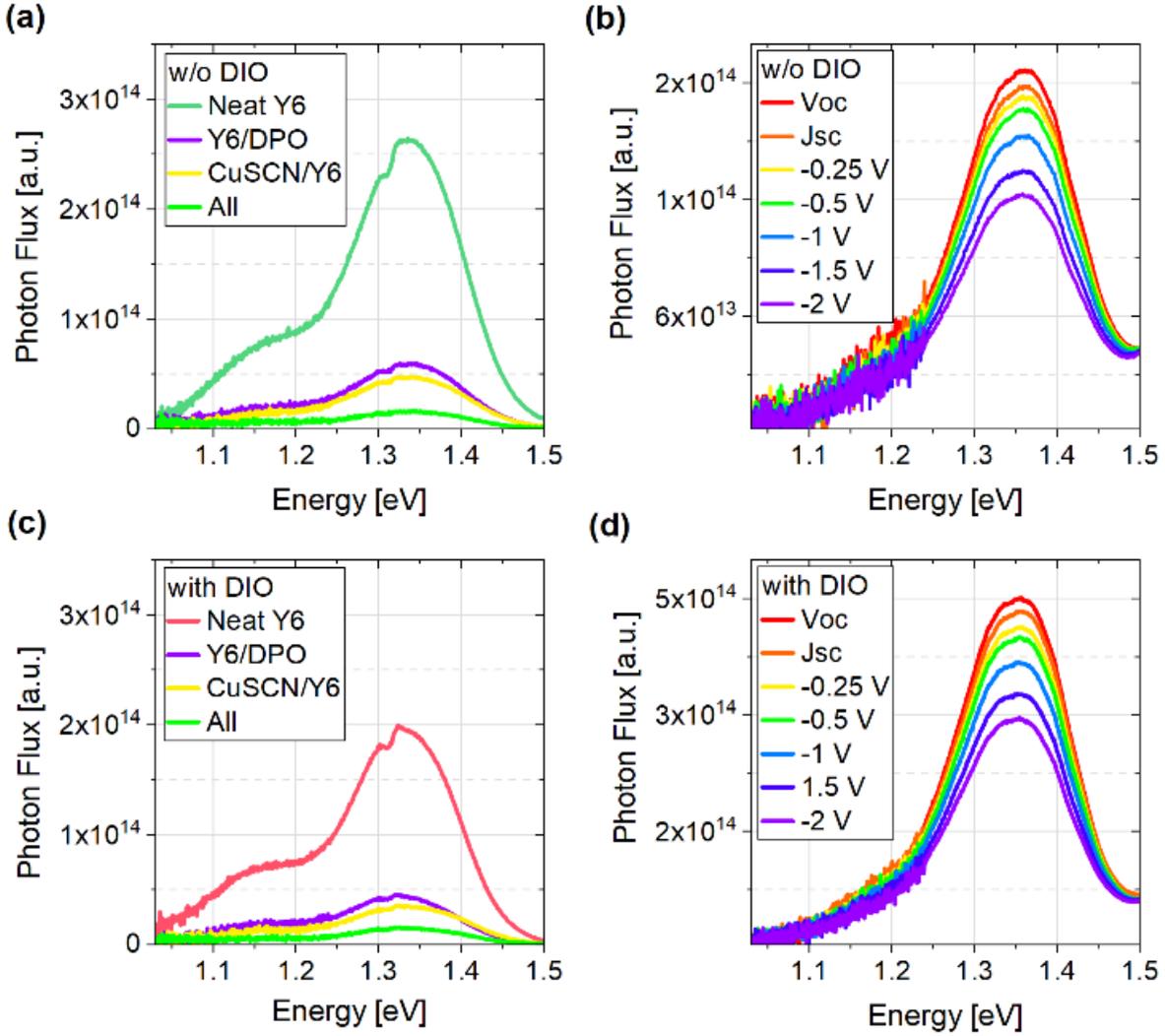

**Figure 3** Photoluminescence spectra of the Y6 films with a 15 nm thickness **a.** without and **c.** with DIO additive showing the PL quenching when in contact with transport layers. Photoluminescence spectra of a device with 15 nm Y6 active layer **b.** with no additive and **d.** with DIO additive in ITO/CuSCN/Y6/Phen-NaDPO/Al structure, at $V_{OC}$, $J_{SC}$ conditions and under several different applied negative bias conditions.

To understand the PL intensity field dependence, we investigate if charge generation is field dependent, utilizing TDCF experiments as a function of electric field. The experimental details on TDCF have been described elsewhere.[28] In short, the device is excited with a short laser pulse (~5 ns) while being held at a given pre-bias ($V_{pre}$). After a delay time of 6 ns (once the laser is switched off) all charges are extracted by applying a high reverse collection bias ($V_{coll}$). To ensure that non-geminate losses are insignificant during the measurement, we apply a sufficiently large $V_{coll}$ of -2 V and the laser intensity is chosen low enough to lie in the linear regime (the extracted charge is strictly proportional to the laser fluence) where second order recombination is negligible.[29] Then, the total extracted charge (Q) is a direct measure of the efficiency of free charge generation under these conditions. **Figure 4a** shows the results of such a measurement for both DIO and no DIO devices (with the transport layers), where $V_{pre}$ is swept from reverse bias to $V_{OC}$. Here, the excitation energy was 2.38 eV. In both cases, we find that the total charge $Q_{tot}$ does not depend on the applied bias $V_{pre}$, even when approaching $V_{OC}$, meaning that the photocurrent does not suffer from increased geminate recombination when decreasing the internal field. Furthermore, such field independent charge generation mechanism suggests that the field dependent PL, is not due to exciton dissociation but rather free carriers; where with increasing reverse bias, extraction is improved and thereby fewer charges are left to recombine through the reformed exciton (reduced PL intensity).

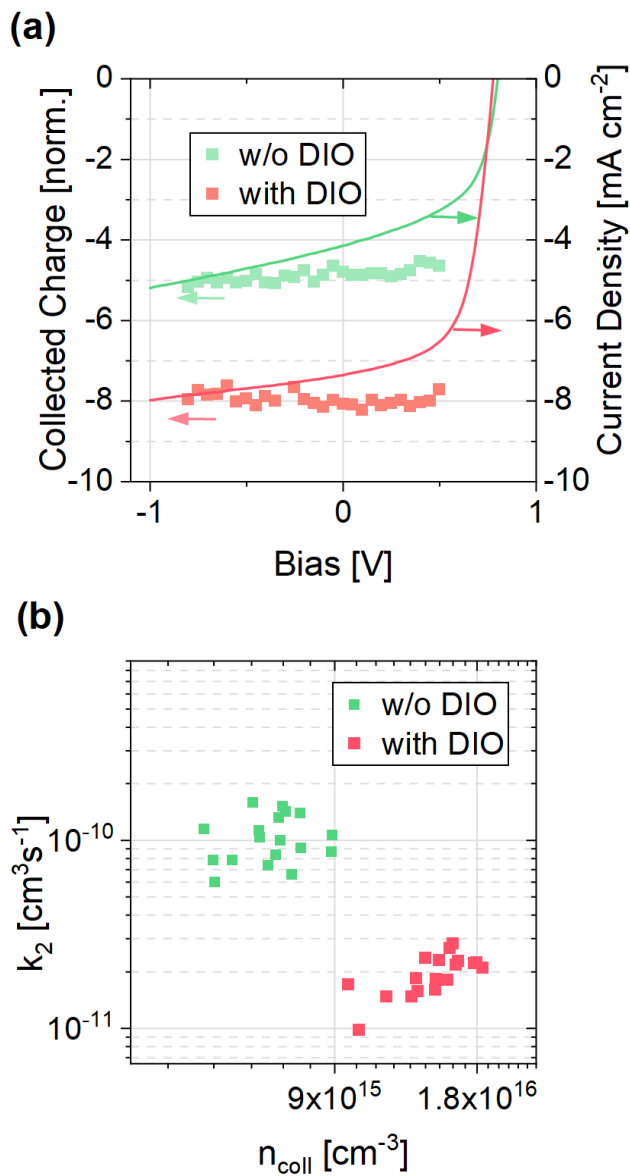

**Figure 4a.** Bias-dependent free charge generation (symbols, left axis) for an inverted Y6 device measured by TDCF for an excitation of 2.07 eV with a low fluence of 0.05 µJ cm−2 and $V_{coll}$= −2 V. For comparison, the current density–voltage characteristics of the device under simulated AM1.5G light is also shown (solid lines, right axis). **b.** Bimolecular recombination coefficient as a function of charge carrier density, obtained from BACE measurements for Y6 with 100 nm thickness without DIO additive (solid pink squares), and with DIO additive (solid green squares)

In blends, it has been argued that CT separation can be assisted by various processes such as driving force,[30] entropy,[31] high local mobilities,[32] and delocalization of charges on aggregated phases of the donor and/or the acceptor.[33,34] In addition, several recent papers highlighted the role of energetic disorder in providing low energy sites for the dissociation of CT states in blends or even singlet excitons in neat organic semiconductors. For instance, Hood and Kassal concluded that a Gaussian disorder $\sigma$ of 100 meV is sufficient to reduce the free-energy barrier to ca. 25 meV, meaning that thermal energy can be sufficient to dissociate CT states at room temperature.[31]

Accordingly, we performed temperature dependent space-charge limited current (SCLC) experiments in electron- and hole-only devices to quantify mobility at room temperature and the energetic disorder in Y6 devices (since this approach has been shown to be sensitive to the shape and width of the DOS) (Figure S6). Energetic disorder obtained using the Gaussian disorder model (GDM) and mobility values are tabulated in Table 2. Whilst the combined disorder is rather large, however other NFAs also exhibit similar or even larger values (see Table S4). Therefore, such disorder values cannot be solely responsible for energetic disorder as the main driving force for charge separation.

| Additive | $\mu_e$[cm²/Vs] | $\sigma_{LUMO}$ [meV] | $\mu_h$[cm²/Vs] | $\sigma_{HOMO}$ [meV] |
|---|---|---|---|---|
| with DIO | $2.1 \times 10^{-3}$ | 70 | $9.6 \times 10^{-4}$ | 76 |
| w/o DIO | $2.4 \times 10^{-3}$ | 71 | $1.8 \times 10^{-4}$ | 74 |

**Table 2.** The mobility values were estimated from the single carrier SCLC devices with 200 nm of active layer thickness, the spatial disorder was calculated with the temperature dependent SCLC mobility for the temperature range of 320-220 K (see SI for more details).

Now turning to investigate non-geminate recombination and to determine the recombination coefficient $k_{rec}$ as a function of carrier density, we employed bias assisted charge extraction (BACE). In BACE, the device is held at $V_{OC}$ under steady-state illumination and as soon as the light is turned off, a high reverse bias is applied to extract all charges. A recombination order close to 2 rules out trap-assisted recombination. The analysis of the recombination data according to $R = k_2 n_{coll}^2$ as shown in **Figure 4b** yields bimolecular recombination coefficient values of 2× 10⁻¹¹ cm³ s⁻¹ and 1 × 10⁻¹⁰ cm³ s⁻¹ for Y6 with and without DIO respectively. We note that these values are up to one order of magnitude higher than what is reported for state of the art PM6:Y6 7× 10⁻¹² cm³ s⁻¹ - 2 × 10⁻¹¹ cm³ s⁻¹.[35, 36]

The efficiency limitation that comes with the high $k_2$ value in the sample without DIO can be illustrated with numerical drift-diffusion simulations. In **Figure 5** we show that by using our experimental data as input to simulation, we fully reproduce the JV curves of the neat Y6 devices with and without DIO, using our measured parameters (a summary of all simulation parameters is given Table S3). Whilst the difference in recombination between the two devices can be explained by the hole mobility difference in improving the competition between extraction vs recombination, however we should not underestimate the role CuSCN plays. In comparison, recombination of free carriers in Y6 on glass (without transport layers), as reported by Hodgkiss and co-workers, is on the timescale of 100 pico-seconds.[12] This insinuates that in the presence of transport layers, the free charges generated in the bulk can travel to the respective transport layer and supress recombination.

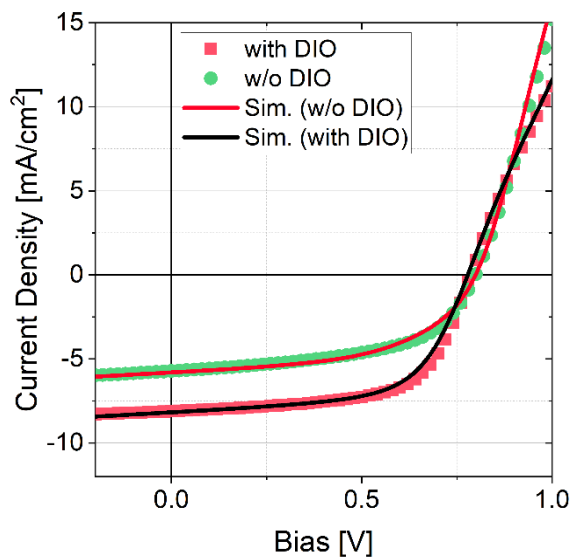

**Figure 5.** Current density-voltage characteristics of the 100 nm Y6 devices with DIO ($V_{OC}$= 0.78 V, $J_{SC}$= 8,1 mA/cm$^2$, *FF*=63 %, PCE=4%) and w/o DIO ($V_{OC}$= 0.80 V, $J_{SC}$=5.7, FF=54%, PCE=2.5 %) additive with structure ITO/CuSCN/Y6/Phen-NaDPO/Al (fully reflecting 100nm) with the device area of 6 mm$^2$ under simulated AM1.5G light. Open squares are the experimental results and the solid lines are the drift-diffusion simulation results generated with using the experimentally determined values for k$_2$ and values close to the measured $\mu_{e/h}$. A shunt resistance ($R_{shunt}$ =$\mathbf{8.6 \times 10^2}$Ωcm$^2$) and series resistance ($R_s$=**1** Ωcm$^2$) were implemented.

By comparing the device performance of Y6 with those of Y5 and eC-9 (see Figure S9 and Table S4 for details) we conclude that our results question charge thermalisation in a broad DOS as the origin of efficient charge generation in our Y6 neat device. Instead, they favour a picture with aggregation-dependent electron affinity; providing an energy cascade to dissociate the excitons and drive electron and holes out from the more disordered domains into the well-crystallized domains of neat Y6. The structure of the core unit of Y6 is quite different from that of traditional acceptor (A) donor (D) acceptor NFAs. [37] Y6 has a unique ADA'DA structure and the core is angular with an electron-deficient benzothiadiazole fragment in the centre. It has four side chains, with two attached at the inside pyrrole rings and two attached at the outside thiophene rings. Due to the steric repulsion between the two inside alkyl chains, the molecular plane of Y6 shows a certain degree of twist, and the inside alkyl chains are out of the plane to mitigate over crystallisation. Furthermore, the two outside alkyl chains restrict the rotation of the 2-(5,6-difluoro-3-oxo-2,3-dihydro-1H-inden-1-ylidene) malononitrile (DFIC) end units, leading to a higher degree of conformational rigidity and uniformity for the Y6 acceptor. This structure results in large quadrupole moment, which causes band bending and may facilitate and contribute to charge separation. [38] Furthermore, in comparison to NFAs reported earlier, Y6 film has a more preferential face-on orientation, and clusters of Y6 are better connected, encouraging more efficient transport. Single crystal analysis also revealed that Y6 undergoes rather unique peaking to form a 3D network for efficient charge transport [39] consistent with our transport data. Coupling charge generation properties with efficient transport mobilities for both holes and electrons in Y6, free carriers are spatially separated from one another in the bulk and extracted into the transport layers; thereby achieving long lived free carriers directly observed by photo-Hall and TDCF measurements.

## Conclusions

Combining photo-Hall measurements and PIA, we showed that Y6 material generates free charges upon light excitation. We have characterized full devices consisting of a single active layer component with a PCE of 4.5%, and showed that the free charge generation is a field independent process. Furthermore, by comparing Y6 devices with and without DIO additive, we explained the increase in the $J_{SC}$ by enhanced charge generation due to more exciton dissociation due to aggregation as well as better extraction of holes due to higher mobility. The former is confirmed with the steady state photoluminescence measurements that show suppression of Y6's emission with DIO additive, as well as the photo-Hall effect and PIA measurements that give a steeper increase in the carrier density up to 1 sun. The improved *FF* of the Y6 devices with the additive is explained by balanced charge carrier mobility. The bimolecular recombination is suppressed in the Y6 devices with DIO, as it appears in the bias assisted charge extraction measurements. In the light of these drastic changes in the device characteristics with the DIO additive, we consider that there is a morphology dependent energy cascade in the Y6 films.

## Experimental

### Sample Preparation

The small acceptor molecule Y6 were purchased from 1-Material Inc. Copper thiocyanate (CuSCN), diethyl sulfide (DES), chloroform (CHCl$_3$), 1,8-Diiodooctane (DIO), 3-[6-(diphenylphosphinyl)-2-naphthalenyl]-1,10-Phenanthroline (Phen-NaDPO), and methanol were purchased from Sigma Aldrich. 1-Chloronaphthalene was purchased from Alfa Aesar. 2,9-Dimethyl-4,7-diphenyl-1,10-phenanthroline (BCP) The devices with a regular configuration were fabricated with a structure of ITO/CuSCN/ Y6/Phen-NaDPO/Al. Patterned ITO (Lumtec & PsiOTech Ltd.) substrates were cleaned in a beaker with Hellmanex at 75°C for 1h, wiped with paper tissues then taken to an ultrasonic bath to proceed to the cleaning with Hellmanex for 20 min., deionized water for 20 min., acetone for 15 mins. and isopropanol for 15 min, followed by microwave plasma treatment (4 min at 200 W). 26 mg/ml CuSCN was dissolved in DES at 60°C for 1h and spin coated onto ITO at 2500 rpm. Y6 solutions were prepared either with pure CHCl$_3$ or 0.5%

ARTICLE

DIO (v/v, DIO/ CHCl3) in 20 mg/ml concentration. 100 nm of the active layer was spin coated onto CuSCN hole transport layer at 1700 rpm. Phen-NaDPO electron transport layer and Al electrode were evaporated under a $10^{-6}$-$10^{-7}$ mbar vacuum. For PL measurements the neat films and films with transport layers were spin coated onto the clean glass substrates.

**Current Voltage Characteristics**

*JV* curves were measured using a Keithley 2400 system in a 2-wire source configuration. Simulated AM1.5G irradiation at 100 mWcm-2 was provided by a filtered Oriel Class AAA Xenon lamp and the intensity was monitored simultaneously with a Si photodiode. The sun simulator is calibrated with a KG5 filtered silicon solar cell (certified by Fraunhofer ISE).

**Time Delayed Collection Field (TDCF) and Bias Assisted Charge Collection (BACE)**

In TDCF, the device was excited with a nano second laser pulse from a diode pumped Q-switched Nd:YAG laser (NT242, EKSPLA) with 5 ns pulse duration at a typical repetition rate of 500 Hz. To compensate for the internal latency of the pulse generator, the laser pulse was delayed and homogeneously scattered in an 85 m long silica fibre (LEONI). The devices were held at different pre-bias ($V_{pre}$) were subsequently switched to a large negative bias ($V_{coll}$) (-2V) to collect the photogenerated charges with the laser pulse. $V_{pre}$ and $V_{coll}$ were set by an Agilent 81150A pulse generator through a home-built amplifier, which was triggered by a fast photodiode (EOT, ET-2030TTL). The current flowing through the device was measured via a 10 Ω resistor in series with the sample and recorded with an oscilloscope (Agilent DSO9104H). Great care was taken to avoid free carrier recombination prior to extraction. Therefore, a fast ramp-up (~2.5 ns) of the bias was applied.

In BACE, the devices were held at steady state conditions by illumination with 1 W, 638 nm and 520nm laser diode (insaneware) with a switch-off time of ~10 ns. The laser diode was operated at 500 Hz with a duty cycle of 50%, such that illumination lasted 1 ms and the diode was switched off for also 1 ms. Right after switching off the laser, a high reverse bias was applied to the sample by the same fast pulse generator (Agilent 81150A) as in TDCF measurements, allowing a fast extraction time of $10^{-20}$ ns. The current transients were measured via a 10 Ω resistor in series with the sample and recorded with an oscilloscope (Agilent DSO9104H).

**Photoluminescence (PL)**

PL emission spectra measurements were recorded with an Andor Solis SR393i-B spectrograph with a silicon detector DU420A-BR-DD and an Indium Gallium Arsenide DU491A-1.7 detector. A calibrated Oriel 63355 lamp was used to correct the spectral response. The film under study was excited with a 520 nm laser (insaneware) under negative bias open-circuit (OC) and short-circuit (SC) conditions. PL spectra were recorded with different gratings with centre wavelengths of 800, 1100, and 1400 nm and merged afterwards.

In absolute PL measurements, the same laser was used together with an optical fibre that goes in to the integrating sphere that holds the sample. Measurements were made with 1 sun intensity. The spectra were recorded by Andor Solis SR393i-B spectrograph connected to the integrating sphere with another optical fibre. A calibrated Oriel 63355 lamp was used to correct the spectral response (same as for EL and PL measurements above), which was shone into the integrating sphere. The spectral photon density was obtained from the corrected detector signal (spectral irradiance) by division through the photon energy ($hv$), and the photon numbers of the excitation and emission were calculated from the numerical integration, using a MatLab code. Great care was taken avoid that the measured PLQY values with the integrating sphere were influenced by waveguided light that is outcoupled through the sides of the substrate, and thereby we taped the four sides of our glass substrates with black tape.

**AC Hall measurements**

Hall measurements [40] were performed in 4-probe configuration with Lake Shore 8400 Hall system. All 4 probe combinations showed linear IV response and ohmic contacts signature. We used AC magnetic field (100 mHz) and a lock-in amplifier to enhance the Hall signal. The current through the samples (in the range 0.1-2 nA) was supported by the current source. The Hall data are fully consistent with 4-probe (4c) conductivity measurements.

**Photo-Hall and Photoconductivity measurements**

To give insight on charge transport under illumination, we further studied thin-film conductivity and Hall effect by illuminating them with 1.7 eV light and intensity close to 1 sun. The photoexcited hole concentration is calculated according to Hall effect mobility (Δn=PhC/(e·μh) due to a better signal-to-noise ratio. The lifetime ($\tau_e$) and diffusion length (*L*) were calculated using the following relations $\tau_e = \frac{n}{G}$ and $L = \sqrt{D\tau}$, where *G* is generation rate and *D* is diffusion constant. Due to the high thickness of the samples (100 nm), we assume the total absorption of the light. The decrease of Hall mobility in the sample with additives is explained by the participation of free holes in charge transport, compensating free electron signal: [41]

$$\mu_H = \frac{(\mu_p^2 p - \mu_n^2 n)}{(\mu_p p + \mu_n n)}$$

**Photo Induced Absorption (PIA)**

In PIA measurements, the photoexcitation of a 405-nm continuous wave laser diode (Spectral Products) is modulated at a frequency of 570 Hz by an optical chopper (Thorlabs MC2000B). In EM measurements, a square voltage with a frequency fixed at 370 Hz and a tuneable amplitude provided by a function generator (Keysight 33210A) is applied to the device. The white light emitted from a tungsten halogen lamp is optically directed into a monochromator (Spectral Products DK240), and the monochromatic light existing from the monochromator is used as the probe light, and focused on the studied device to overlap with the photoexcitation light. The change in the transmitted probe light $\Delta T$ induced by the photoexcitation in PIA and the dark injection in EMIA is recorded by a Si photodiode (Thorlabs) and a lock-in amplifier (SR830) and then corrected by a background subtraction. The transmitted probe light T through the unexcited sample is measured by using another optical chopper (Thorlabs MC2000B) to modulate the probe light, the same photodiodes and the lock-in amplifier. Our PIA system has a sensitivity on the order of $10^{-7}$.

**Space charge-limited currents (SCLC)**

Electron-only devices were prepared in ITO/ZnO/Y6(170-200 nm)/Phen-NaDPO (10 nm)/Al (100 nm) configuration. ZnO nanoparticle dispersion in isopropanol (Avantama N-10) was filtered with a 0.45 µm polytetrafluoroethylene filter and spin coated onto ITO at 5000 rpm for 40 s in air and annealed at 120 °C for 30 min. Hole-only devices with the configuration ITO/MoO$_3$ (8 nm)/Y6(170-200 nm)/MoO$_3$ (8 nm)/Ag (100 nm) were prepared by evaporating MoO$_3$ on top of ITO. Then, the active layer was prepared as for solar cell devices, followed by evaporation of 8 nm of MoO$_3$ under a $10^{-6}$-$10^{-7}$ mbar vacuum.

For temperature-dependent measurements, the devices were loaded into a liquid nitrogen-cooled cryostat (VPF-100 Janis) and the temperature was adjusted in a range of 220 K to 320 K using a temperature controller (Lakeshore 335). *JV* data were measured using a Keithley 2400 Source Meter in a two-wire configuration.

**External quantum efficiency (EQE$_{PV}$) and Absorbance**

EQE$_{PV}$ was measured using a home-made setup containing a quartz tungsten halogen lamp, a Thorlabs MC2000B optical chopper at a frequency of 165 Hz, a Bentham TMC300 monochromator, a lock-in amplifier (SR830) and a preamplifier (SR570). The system was calibrated using a standard silicon detector from Enlitech. For EQE measurements as function of temperature, the sample was placed in the same cryostat as *JV* measurements and the modulated light from monochromator was focused on the active area of the sample. The electrical signal of sample was sent to the preamplifier then the lock-in amplifier. The voltage was applied to the sample from the preamplifier. Absorbance was measured with a Varian Cary 5000 in transmission and reflection model with an integrating sphere.

**Spatially and Time Resolved Fluorescence Microscopy**

A 150 fs-pulsed Ti:Sapphire laser with a repetition rate of 76 MHz and wavelength 800 nm is used as excitation source. A telescope with a pinhole spatially cleans the beam, while prism compressors compensate for temporal dispersion. The beam is scanned over the sample plane due to a galvo-mirror followed by a second telescope. A Nikon Plan APO λ NA 1.40 60x oil immersion objective focused the beam to a spot of FWHM of 390 nm. In the detection part, fluorescence is confocally collected by a Single Photon Avalanche Diode (SPAD) from the MPD series (instrument response function of 50 ps) in reflection through an 830 LP filter and a 15 $\mu$m pinhole corresponding to a central spot size of 250 nm diameter in the sample plane. To acquire 2D spatio-temporal maps, the excitation spot is moved away from the detection spot of a well-known distance. Excitons need a certain time to travel the distance between the generation and the collection spot. This results in a time delay which can be identified in an effective increase of the lifetime of fluorescence. The collected fluorescence intensity *I*(x,t) is then integrated every 20 ps at each scanning position (every 100 nm) along the cross-section of the excitation profile.

## Conflicts of interest

There are no conflicts to declare.

## Acknowledgements

This work was supported in part by the Alexander von Humboldt Foundation and Deutsche Forschungsgemeinschaft (DFG, German Research Foundation, through the project Fabulous (NE 410/20, SH 1669/1-1) and the SFB HIOS (Projektnummer 182087777 - SFB 951). A.M. acknowledges financial support from the German Science Foundation (DFG) SPP 2196 and Horizon Europe Framework Programme, call - HORIZON-MSCA-2021-PF-01, acronym - HyPerGreen, agreement number - 101061809. GLGM and NFvH acknowledge support through the MCIN/AEI project PRE2019-091051, the "Severo Ochoa" program for Centers of Excellence in R&D CEX2019-000910-S, Fundació Privada Cellex, Fundació Privada Mir-Puig, and the Generalitat de Catalunya through the CERCA program.

## References

1. L. Zhan, S. Li, X. Xia, Y. Li, X. Lu, L. Zuo, M. Shi, H. Chen, *Adv. Mater.* 2021, **33**, 1.
2. Z.C. Wen, H. Yin, X.T. Hao, *Surfaces and Interfaces* 2021, **23**, 100921.
3. X. Ma, A. Zeng, J. Gao, Z. Hu, C. Xu, J.H. Son, S.Y. Jeong, C. Zhang, M. Li, K. Wang, H. Yan, Z. Ma, Y. Wang, H.Y. Woo, F. Zhang, *Natl. Sci. Rev.* 2021, **8**.
4. L. Perdigón-Toro, H. Zhang, A. Markina, J. Yuan, S.M. Hosseini, C.M. Wolff, G. Zuo, M. Stolterfoht, Y. Zou, F. Gao, D. Andrienko, S. Shoaee, D. Neher, *Adv. Mater.* 2020, **32**.
5. L. Zhu, M. Zhang, J. Xu, C. Li, J. Yan, G. Zhou, W. Zhong, T. Hao, J. Song, X. Xue, Z. Zhou, R. Zeng, H. Zhu, C.C. Chen, R.C.I. MacKenzie, Y. Zou, J. Nelson, Y. Zhang, Y. Sun, F. Liu, *Nat. Mater.* 2022, **21**, 656.
6. S. Chen, Y. Wang, L. Zhang, J. Zhao, Y. Chen, D. Zhu, H. Yao, G. Zhang, W. Ma, R.H. Friend, P.C.Y. Chow, F. Gao, H. Yan, *Adv. Mater.* 2018, **30**.
7. J. Liu, S. Chen, D. Qian, B. Gautam, G. Yang, J. Zhao, J. Bergqvist, F. Zhang, W. Ma, H. Ade, O. Inganäs, K. Gundogdu, F. Gao, H. Yan, *Nat. Energy* 2016, **1**.
8. V.I. Arkhipov, H. Bässler, *Phys. Status Solidi Appl. Res.* 2004, **201**, 1152.
9. S. Karuthedath, J. Gorenflot, Y. Firdaus, N. Chaturvedi, C.S.P. De Castro, G.T. Harrison, J.I. Khan, A. Markina, A.H. Balawi, T.A. Dela Peña, W. Liu, R.Z. Liang, A. Sharma, S.H.K. Paleti, W. Zhang, Y. Lin, E. Alarousu, D.H. Anjum, P.M. Beaujuge, S. De Wolf, I. McCulloch, T.D. Anthopoulos, D. Baran, D. Andrienko, F. Laquai, *Nat. Mater.* 2021, **20**, 378.
10. L. Zhu, Y. Yi, Z. Wei, *J. Phys. Chem. C* 2018, **122**, 22309.
11. K. Nakano, Y. Chen, B. Xiao, W. Han, J. Huang, H. Yoshida, E. Zhou, K. Tajima, *Nat. Commun.* 2019, **10**, 1.
12. M.B. Price, P.A. Hume, A. Ilina, I. Wagner, R.R. Tamming, K.E. Thorn, W. Jiao, A. Goldingay, P.J. Conaghan, G. Lakhwani, N.J.L.K. Davis, Y. Wang, P. Xue, H. Lu, K. Chen, X. Zhan, J.M. Hodgkiss, *Nat. Commun.* 2022, **13**, 1.
13. R. Wang, C. Zhang, Q. Li, Z. Zhang, X. Wang, M. Xiao, *J. Am. Chem. Soc.* 2020, **142**, 12751.
14. Y. Firdaus, V.M. Le Corre, S. Karuthedath, W. Liu, A. Markina, W. Huang, S. Chattopadhyay, M.M. Nahid, M.I. Nugraha, Y. Lin, A. Seitkhan, A. Basu, W. Zhang, I. McCulloch, H. Ade, J. Labram, F. Laquai, D. Andrienko, L.J.A. Koster, T.D. Anthopoulos, Nat. Commun. 2020, **11**, 5220.
15. L. Zhu, J. Zhang, Y. Guo, C. Yang, Y. Yi, Z. Wei, *Angew. Chemie - Int. Ed.* 2021, **60**, 15348.
16. D. Neusser, B. Sun, W.L. Tan, L. Thomsen, T. Schultz, L. Perdigón-Toro, N. Koch, S. Shoaee, C.R. McNeill, D. Neher, S. Ludwigs, *J. Mater. Chem. C* 2022, **10**.
17. J. Wu, X. Jiang, X. Peng, *ACS Appl. Energy Mater.* 2022, **5**, 11646.
18. T.L. Nguyen, T.H. Lee, B. Gautam, S.Y. Park, K. Gundogdu, J.Y. Kim, H.Y. Woo, *Adv. Funct. Mater.* 2017, **27**.
19. Y. He, N. Li, C.J. Brabec, *Org. Mater.* 2021, **03**, 228.
20. H. Yu, Z. Qi, J. Zhang, Z. Wang, R. Sun, Y. Chang, H. Sun, W. Zhou, J. Min, H. Ade, H. Yan, *J. Mater. Chem. A* 2020, **8**, 23756.
21. N. Wijeyasinghe, F. Eisner, L. Tsetseris, Y.H. Lin, A. Seitkhan, J. Li, F. Yan, O. Solomeshch, N. Tessler, P. Patsalas, T.D. Anthopoulos, *Adv. Funct. Mater.* 2018, **28**.
22. X. Xu, Y. Qi, X. Luo, X. Xia, X. Lu, J. Yuan, Y. Zhou, Y. Zou, *Fundam. Res.* 2022.
23. A. Seitkhan, M. Neophytou, M. Kirkus, E. Abou-Hamad, M.N. Hedhili, E. Yengel, Y. Firdaus, H. Faber, Y. Lin, L. Tsetseris, I. McCulloch, T.D. Anthopoulos, *Adv. Funct. Mater.* 2019, **29**.
24. L.Q. Phuong, S.M. Hosseini, C.W. Koh, H.Y. Woo, S. Shoaee, *J. Phys. Chem. C* 2019, **123**, 27417.
25. Q. Wei, J. Yuan, Y. Yi, C. Zhang, Y. Zou, *Natl. Sci. Rev.* 2021, **8**, 10.
26. T. Xiao, H. Xu, G. Grancini, J. Mai, A. Petrozza, U.S. Jeng, Y. Wang, X. Xin, Y. Lu, N.S. Choon, H. Xiao, B.S. Ong, X. Lu, N. Zhao, *Sci. Rep.* 2014, **4**.
27. L. Zhu, Z. Tu, Y. Yi, Z. Wei, *J. Phys. Chem. Lett.* 2019, **10**, 4888.
28. J. Kniepert, I. Lange, N.J. Van Der Kaap, L.J.A. Koster, D. Neher, *Adv. Energy Mater.* 2014, **4**.
29. J. Kurpiers, T. Ferron, S. Roland, M. Jakoby, T. Thiede, F. Jaiser, S. Albrecht, S. Janietz, B.A. Collins, I.A. Howard, D. Neher, *Nat. Commun.* 2018, **9**, 1.
30. S. Shoaee, *J. Photonics Energy* 2012, **2**, 021001.
31. D. Balzer, T.J.A.M. Smolders, D. Blyth, S.N. Hood, I. Kassal, *Chem. Sci.* 2021, **12**, 2276.
32. S. Shoaee, A. Armin, M. Stolterfoht, S.M. Hosseini, J. Kurpiers, D. Neher, *Sol. RRL* 2019, **1900184**, 1900184.
33. A.E. Jailaubekov, A.P. Willard, J.R. Tritsch, W.L. Chan, N. Sai, R. Gearba, L.G. Kaake, K.J. Williams, K. Leung, P.J. Rossky, X.Y. Zhu, *Nat. Mater.* 2013, **12**, 66.
34. D. Balzer, I. Kassal, *Sci. Adv.* 2022, **8**, 1.
35. A. Karki, J. Vollbrecht, A.L. Dixon, N. Schopp, M. Schrock, G.N.M. Reddy, T. Nguyen, *Adv. Mater.* 2019, **31**, 1903868.
36. S.M. Hosseini, N. Tokmoldin, Y.W. Lee, Y. Zou, H.Y. Woo, D. Neher, S. Shoaee, *Sol. RRL* 2020, **2000498**, 1.
37. C. Yan, S. Barlow, Z. Wang, H. Yan, A.K.Y. Jen, S.R. Marder, X. Zhan, *Nat. Rev. Mater.* 2018, **3**, 1.
38. M. Schwarze, K.S. Schellhammer, K. Ortstein, J. Benduhn, C. Gaul, A. Hinderhofer, L. Perdigón Toro, R. Scholz, J. Kublitski, S. Roland, M. Lau, C. Poelking, D. Andrienko, G. Cuniberti, F. Schreiber, D. Neher, K. Vandewal, F. Ortmann, K. Leo, *Nat. Commun.* 2019, **10**, 1.
39. G. Zhang, X.K. Chen, J. Xiao, P.C.Y. Chow, M. Ren, G. Kupgan, X. Jiao, C.C.S. Chan, X. Du, R. Xia, Z. Chen, J. Yuan, Y. Zhang, S. Zhang, Y. Liu, Y. Zou, H. Yan, K.S. Wong, V. Coropceanu, N. Li,

C.J. Brabec, J.L. Bredas, H.L. Yip, Y. Cao, Nat. Commun. 2020, **11**, 3943

40　F. Peña-Camargo, J. Thiesbrummel, H. Hempel, A. Musiienko, V.M. Le Corre, J. Diekmann, J. Warby, T. Unold, F. Lang, D. Neher, M. Stolterfoht, *Appl. Phys. Rev.* 2022, **9**, 021409.

41　A. Musiienko, R. Grill, P. Moravec, P. Fochuk, I. Vasylchenko, H. Elhadidy, L. Šedivý, *Phys. Rev. Appl.* 2018, **10**, 1.